\documentclass[a4paper,11pt]{article}
\pdfoutput=1 

\usepackage{jcappub} 

\usepackage[T1]{fontenc} 
\usepackage{caption}

\newcommand{\beq}{\begin{equation}}
\newcommand{\eeq}{\end{equation}}
\newcommand{\bea}{\begin{eqnarray}}
\newcommand{\eea}{\end{eqnarray}}
\newcommand{\beqn}{\begin{equation*}}
\newcommand{\eeqn}{\end{equation*}}
\newcommand{\bean}{\begin{eqnarray*}}
\newcommand{\eean}{\end{eqnarray*}}

\newcommand*{\cref}[1]{Chapter~\ref{#1}}

\title{\boldmath  Two-parameter continuous deformation of Starobinsky inflation  as a bridge between Planck and ACT DESI data with $N_\star\in(50,60)$}

\author[a]{Gabriel Germ\'an}
\affiliation[a]{Instituto de Ciencias F\'{\i}sicas, Universidad Nacional Aut\'onoma de M\'exico, Av. Universidad s/n, Cuernavaca, Morelos 62210, Mexico}
\emailAdd{gabriel@icf.unam.mx}

\abstract{%
We present a family of plateau-type inflationary potentials, eq.~\eqref{Vgeneral}, and analyze a two-parameter $\alpha\beta$-Starobinsky specialization that interpolates continuously between a \emph{maximal} plateau ($V\!\to\!V_0$) and a \emph{submaximal} plateau ($V\!\to\!V_\infty<V_0$), where $V_0$ is the overall scale of the potential. In the $\beta\!\to\!0$ limit our framework reproduces the $\alpha$-Starobinsky predictions $n_s\simeq 1-2/N_\star$ and $r\simeq 12\alpha/N_\star^2$. For $\beta>0$ with $x_\star\gg 1/\beta$ the slow-roll scaling laws change to $n_s\simeq 1-\frac{4}{3N_\star},\, r\simeq\mathcal{C}(\alpha,\beta)\,N_\star^{-4/3},$ with an explicit coefficient $\mathcal{C}(\alpha,\beta)$ set by the plateau truncation. This deformation lifts $n_s$ at fixed $N_\star$ while further suppressing $r$, reconciling the Planck~2018 constraint $n_s=0.9649\pm0.0042$ (68\% CL) and BICEP/Keck18 data  $r_{0.05}<0.036$ (95\% CL), with the higher central values $n_s\sim0.97$--$0.98$ preferred by ACT+DESI~DR2 (BAO), within the theoretically motivated interval $N_\star\in(50,60)$ and without exotic reheating. We provide an exact identity for $V/V'$ enabling analytic control of $N_\star$, a practical crossover criterion $\beta\,x_\star\ll1$ vs.\ $\gg1$, and a transparent mapping between $(\alpha,\beta)$ and the observables $(n_s,r,N_\star)$. These yield sharp, testable signatures, particularly the softened $N_\star$-scaling of $r$, that distinguish a maximal from a submaximal plateau with upcoming CMB and LSS data.}

\begin{document}
\maketitle
\flushbottom

\section{Introduction}\label{Intro}

Precision measurements of the cosmic microwave background (CMB) have established a red-tilted, nearly scale-invariant spectrum with tight limits on primordial tensors. In particular, Planck~2018 determines the scalar spectral index to be $n_s=0.9649\pm0.0042$ (68\% CL)   \cite{Planck:2018jri} and  BICEP/Keck Array BK18, places a 95\% CL upper limit $r_{0.05}<0.036$ \cite{Tristram:2021tvh}. These results favor single-field slow-roll inflation  with concave potentials and broadly motivate plateau-type dynamics (for reviews on inflation see e.g., \cite{Linde:1984ir}-\cite{Martin:2013tda}). At the same time, combined analyses including ACT polarization and DESI~DR2 BAO indicate slightly higher central values of $n_s$ (around $0.97\!-\!0.98$)  \cite{ACT:2025fju}-\cite{DESI:2025zgx} tightening the phenomenological target while keeping $r$ small. Achieving such $n_s$ within the theoretically preferred window $N_\star\in(50,60)$ can be challenging for well-known models\footnote{In \cite{Kallosh:2025rni}-\cite{Odintsov:2025wai} we refer to recent articles that propose to reconcile inflationary models with the latest CMB data.}, including the original Starobinsky model \cite{Starobinsky:1980te} and its $\alpha$-generalizations \cite{SPT-3G:2025bzu}-\cite{German:2025ide}, which often require either larger $N_\star$ or nonstandard reheating  to raise $n_s$ appreciably (for reviews on reheating see e.g., \cite{Bassett:2005xm}, \cite{Allahverdi:2010xz}, \cite{Amin:2014eta}).

To address this tension while preserving slow-roll robustness, we consider a minimal but flexible family of plateau-type potentials
\begin{equation}
\label{Vgeneral}
  V(\phi)=V_0\!\left(1-e^{-\frac{(\phi/M_{\rm Pl})^p}{\,c_1+c_2(\phi/M_{\rm Pl})^q\,}}\right)^{m},
\end{equation}
where $M_{\rm Pl}=2.435\times10^{18}\,\mathrm{GeV}$ is the reduced Planck mass. Setting $x\equiv\phi/M_{\rm Pl}$ and
\begin{equation}
\label{Edef}
  \mathcal{E}(x)\equiv\frac{x^p}{c_1+c_2\,x^q}
  \qquad\Rightarrow\qquad
  V(x)=V_0\!\left(1-e^{-\mathcal{E}(x)}\right)^{m},
\end{equation}
with $c_1,c_2$ real constants for each member of the family specified by positive parameters $\{m,p,q\}$, the large-field behavior
\begin{equation}
\label{Easympt}
  \mathcal{E}(x)\simeq
  \begin{cases}
    \dfrac{1}{c_2}\,x^{p-q}, & c_2>0,\\[6pt]
    \dfrac{x^p}{c_1},         & c_2=0,
  \end{cases}
\end{equation}
organizes three qualitatively distinct regimes:
\begin{itemize}
\item[(a)] If $p>q$ (or $c_2=0$), then $\mathcal{E}\to\infty$ and $V\to V_0$ which defines a \emph{maximal} concave plateau with attractor-type predictions $n_s\simeq1-2/N_\star$ and $r\propto N_\star^{-2}$ after an effective field redefinition. Here $N_\star$ is the number of $e$-folds from the pivot scale $k_*$ to the end of inflation.
\item[(b)] If $p=q$ with $c_2>0$, then $\mathcal{E}\to1/c_2$ and $V\to V_0\!\left(1-e^{-1/c_2}\right)^m<V_0$ defines  a \emph{submaximal} concave plateau, typically yielding very small $r$ when the approach to the plateau is sufficiently fast.
\item[(c)] If $p<q$, then $\mathcal{E}\to0$ and $V\to0$ there is no high plateau; the potential flattens to zero at large field, often producing even smaller $r$ with greater sensitivity to reheating details.
\end{itemize}
For the large-field case $p>q$, the substitution $y\equiv x^{p-q}$ (or $y\equiv x^p$ if $c_2=0$) yields near the plateau
\begin{equation}
\label{VplateauE}
  V(\phi)\approx V_0\!\left(1-e^{-y/\tilde{c}}\right)^{m},
  \qquad
  \tilde{c}=
  \begin{cases}
    c_2, & (p>q),\\
    c_1, & (c_2=0),
  \end{cases}
\end{equation}
which, after a canonical field rescaling, behaves as an $E$-model in the $\alpha$-attractor class with
\begin{equation}
\label{alphaAttractorScaling}
  n_s\simeq1-\frac{2}{N_\star},\qquad
  r\simeq\frac{12\,\alpha_{\rm eff}}{N_\star^{2}},
\end{equation}
where $\alpha_{\rm eff}$ depends on $m$, the growth rate set by $p-q$, and $\tilde{c}$. This explains why broad regions with $p>q$ and $m\in\{1,2\}$ reproduce the concave-plateau phenomenology favored by Planck and ACT.

Within this general framework, we focus on the specialization $\{m,p,q\}=\{2,1,1\}$ with the reparametrization
\begin{equation}
\label{cmap}
  c_1=\sqrt{\tfrac{3\alpha}{2}},\qquad c_2=\beta\sqrt{\tfrac{3\alpha}{2}},
\end{equation}
which yields what we call the \emph{$\alpha\beta$-Starobinsky} potential
\begin{equation}
\label{ValphaBeta}
  V(\phi)=V_0\!\left(1-\exp\!\left[
    -\frac{\sqrt{2/(3\alpha)}\,x}{1+\beta x}
  \right]\right)^{2}, \qquad x\equiv\frac{\phi}{M_{\rm Pl}}.
\end{equation}
For $\beta=0$ this reduces to the $\alpha$-Starobinsky (E-model) potential  \cite{Ellis:2013nxa},  \cite{Kallosh:2013hoa}, \cite{Kallosh:2013yoa}, \cite{Ellis:2019bmm}, and for $\beta=0$ with $\alpha=1$ to the original Starobinsky potential in the Einstein frame \cite{Starobinsky:1980te}. For $\beta>0$ the asymptotic plateau is truncated to a submaximal height $V_\infty<V_0$ and the approach to the plateau changes from exponential to power-law, modifying the slow-roll scalings that determine $(n_s,r)$.

\subsection{Main results and outline}\label{sec:outline}

Using the exact identity
\begin{equation}
\label{VoverVprime-intro}
  \frac{V}{V'}=\frac{M_{\rm Pl}}{2}\sqrt{\frac{3\alpha}{2}}\,
  \frac{1-E(x)}{E(x)}\,(1+\beta x)^{2},
  \qquad
  E(x)\equiv\exp\!\left(-\frac{\sqrt{2/(3\alpha)}\,x}{1+\beta x}\right),
\end{equation}
we compute $N_\star$ exactly via a change of variables and identify two predictive regimes controlled by the criterion $\beta x_\star\sim1$, where $x_\star\equiv\phi_\star/M_{\rm Pl}$ at horizon crossing
\begin{equation}
\label{regime-alpha-intro}
  \alpha\text{-Starobinsky-like}\;(\beta x_\star\ll1):\quad
  n_s\simeq1-\frac{2}{N_\star},\quad r\simeq\frac{12\alpha}{N_\star^2},
\end{equation}
\begin{equation}
\label{regime-alphabeta-intro}
  \alpha\beta\text{-Starobinsky}\;(\beta x_\star\gg1):\quad
  n_s\simeq1-\frac{4}{3N_\star},\quad
  r\simeq\mathcal{C}(\alpha,\beta)\,N_\star^{-4/3}.
\end{equation}
The deformation lifts $n_s$ at fixed $N_\star$ and suppresses $r$, softening its scaling from $N_\star^{-2}$ to $N_\star^{-4/3}$. For the pivot $k_\ast=0.05\,\mathrm{Mpc}^{-1}$, this enables agreement with Planck~2018 bounds and accommodation of the slightly higher $n_s$ preferred by ACT+DESI~DR2 (BAO) within $N_\star\in(50,60)$, without super-60 $e$-folds or exotic post-inflationary histories. Fixing $V_0$ via $A_s\simeq2.1\times10^{-9}$, $x_{\rm end}$ from $\epsilon_V=1$, and verifying small spectral running, we also find the deformation reduces the field excursion $\Delta\phi$ while preserving monotonicity and concavity. The rest of the paper develops these results in full, providing closed-form expressions connecting $(\alpha,\beta)$ to $(n_s,r,N_\star)$ and enabling straightforward parameter inference from updated measurements and reheating priors.

\section{The $\alpha\beta$-Starobinsky model: $\{m,p,q\}=\{2,1,1\}$}\label{particularcase}

The specialization $\{m,p,q\}=\{2,1,1\}$ of eq.~\eqref{Vgeneral} gives
\begin{equation}
\label{Vsimple}
  V(\phi)=V_0\!\left(1-e^{-\frac{(\phi/M_{\rm Pl})}{\,c_1+c_2(\phi/M_{\rm Pl})\,}}\right)^{2},
\end{equation}
where $c_1$ and $c_2$ control, respectively, the curvature near the minimum and the asymptotic plateau height.

\noindent\emph{Width near the minimum.}
For $|x|\ll1$,
\begin{equation}
\label{Esmallx}
  \mathcal{E}(x)=\frac{x}{c_1+c_2 x}
  =\frac{x}{c_1}-\frac{c_2}{c_1^{2}}x^{2}+\mathcal{O}(x^{3}),
\end{equation}
so that
\begin{equation}
\label{Vsmallx}
  V(\phi)=V_0\,\frac{\phi^{2}}{c_1^{2}M_{\rm Pl}^{2}}
  +\mathcal{O}\!\left(\frac{\phi^{3}}{M_{\rm Pl}^{3}}\right),
  \qquad V''(0)=\frac{2V_0}{c_1^{2}M_{\rm Pl}^{2}}.
\end{equation}
Smaller $c_1$ yields a narrower potential (larger curvature) near the minimum, while larger $c_1$ makes it wider.

\noindent\emph{Asymptotic height.}
For $\phi\to\infty$ (with $c_2>0$),
\begin{equation}
\label{Vheight}
  V(\phi)\longrightarrow V_0\!\left(1-e^{-1/c_2}\right)^{2}<V_0,
\end{equation}
so $c_2$ controls the plateau height. Since the horizon-crossing field value $\phi_k$ typically lies closer to the plateau than to the minimum, observables at $k$ are mostly sensitive to $c_2$, with $c_1$ acting as a mild modulation through the potential width. As shown below, mean values of $n_s$ consistent with Planck~2018 and ACT+DESI+BAO can be achieved while keeping $N_\star\in(50,60)$, and the asymptotic behavior of $(n_s,r)$ can deviate from the universal $\alpha$-attractor scalings depending on $c_2$.

\subsection{Limiting cases and exponent structure}\label{StaroGene}

With $c_1$ and $c_2$ as in eq.~\eqref{cmap}, eq.~\eqref{Vsimple} reduces to the $\alpha\beta$-Starobinsky potential eq.~\eqref{ValphaBeta}. Introducing $a\equiv\sqrt{2/(3\alpha)}$, the two notable limiting cases are
\begin{align}
  \beta=0:&\quad V(\phi)=V_0\!\left(1-e^{-ax}\right)^{2}
    \quad(\alpha\text{-attractor }E\text{-model}),\label{ValphaModel}\\
  \beta=0,\;\alpha=1:&\quad V(\phi)=V_0\!\left(1-e^{-\sqrt{2/3}\,x}\right)^{2}
    \quad(\text{Starobinsky potential}).\label{VStaro}
\end{align}
For $\beta>0$, writing the exponent as
\begin{equation}
\label{ydef}
  y(x)\equiv-\frac{ax}{1+\beta x},
\end{equation}
one sees that $y(x)\to-a/\beta$ as $x\to\infty$, so the plateau is truncated to a finite, submaximal height (see Sec.~\ref{Asymptotics}).

\subsection{Asymptotic properties for $\beta>0$}\label{Asymptotics}

As $x\to\infty$,
\begin{equation}
\label{asymE}
  \frac{x}{1+\beta x}\to\frac{1}{\beta},\qquad
  y(x)\to-\frac{a}{\beta}\equiv-k,\qquad
  E_\infty\equiv e^{-k},
\end{equation}
giving
\begin{equation}
\label{Vinf}
  V_\infty\equiv\lim_{x\to\infty}V(x)=V_0\bigl(1-e^{-k}\bigr)^{2}<V_0
  \qquad(\beta>0).
\end{equation}
Expanding for $x\gg1/\beta$ via $x/(1+\beta x)=(1/\beta)\bigl(1-1/(\beta x)+\cdots\bigr)$, one finds
\begin{equation}
\label{Vtail}
  V(x)=V_\infty
  -\frac{2V_0\,e^{-k}\bigl(1-e^{-k}\bigr)\,k}{\beta\,x}
  +\mathcal{O}\!\left(\frac{1}{x^{2}}\right),
\end{equation}
so the approach to the plateau is power-law ($\propto x^{-1}$) rather than exponential as in the $\beta=0$ case.

\subsection{Slow-roll parameters, $V/V'$, and $N_\star$}\label{SR}

Define
\begin{equation}
\label{Eofx}
  E(x)\equiv e^{y(x)}=\exp\!\left(-\frac{ax}{1+\beta x}\right).
\end{equation}
The slow-roll parameters are
\begin{equation}
\label{VprimeOverV}
  \frac{V'}{V}=\frac{2a}{M_{\rm Pl}}\,
  \frac{E}{1-E}\,\frac{1}{(1+\beta x)^{2}},
\end{equation}
\begin{equation}
\label{epsilonV}
  \epsilon\equiv\frac{M_{\rm Pl}^{2}}{2}\!\left(\frac{V'}{V}\right)^{2}
  =\frac{4}{3\alpha}\left(\frac{E}{1-E}\right)^{2}\frac{1}{(1+\beta x)^{4}},
\end{equation}
\begin{equation}
\label{etaV}
  \eta\equiv M_{\rm Pl}^{2}\frac{V''}{V}
  =M_{\rm Pl}^{2}\frac{d}{d\phi}\!\left(\frac{V'}{V}\right)+2\epsilon.
\end{equation}
The key quantity for the $e$-fold integral is
\begin{equation}
\label{VoverVprime}
  \frac{V}{V'}=\frac{M_{\rm Pl}}{2a}\,\frac{1-E}{E}\,(1+\beta x)^{2},
\end{equation}
giving
\begin{equation}
\label{Nstarx}
  N_\star=\frac{1}{2a}\int_{x_{\rm end}}^{x_\star}
  \frac{1-E(x)}{E(x)}\,(1+\beta x)^{2}\,dx.
\end{equation}
The exact change of variable $y=-ax/(1+\beta x)$ transforms this into
\begin{equation}
\label{Nstary}
  N_\star=\frac{a^{2}}{2}\int_{y_\star}^{y_{\rm end}}
  \frac{1-e^{y}}{e^{y}}\;\frac{dy}{\bigl(a+\beta y\bigr)^{4}},
\end{equation}
which is well-suited for systematic asymptotic expansion (e.g., around $y=-k$ in the submaximal regime). For $\beta>0$ and $x_\star\gg1/\beta$, setting $E(x)\approx E_\infty=e^{-k}$ constant yields
\begin{equation}
\label{Nasym}
  N_\star\simeq\frac{1}{6a\beta}\left(\frac{1-E_\infty}{E_\infty}\right)\,(1+\beta x_\star)^{3},
\end{equation}
\begin{equation}
\label{epsAsym}
  \epsilon_\star\simeq\frac{4}{3\alpha}
  \left(\frac{E_\infty}{1-E_\infty}\right)^{2}(1+\beta x_\star)^{-4},
\end{equation}
from which the submaximal-regime scalings of Sec.~\ref{Regimes} follow by eliminating $x_\star$ (see Appendix).

\section{Phenomenological regimes and continuous crossover}\label{Regimes}

\paragraph{(i) $\alpha$-Starobinsky regime ($\beta=0$ or $\beta x_\star\ll1$).}
The standard attractor results are recovered  \cite{Kallosh:2013hoa}, \cite{Kallosh:2013yoa}
\begin{equation}
\label{alphaRegime}
  n_s\simeq1-\frac{2}{N_\star},\qquad r\simeq\frac{12\,\alpha}{N_\star^{2}}.
\end{equation}
For $\alpha=1$ (Starobinsky), $r\simeq12/N_\star^{2}$.

\paragraph{(ii) $\alpha\beta$-Starobinsky regime ($\beta>0$, $x_\star\gg1/\beta$).}
Using eqs.~\eqref{Nasym}--\eqref{epsAsym}, one has $N_\star\propto(1+\beta x_\star)^3$ and $\epsilon_V\propto(1+\beta x_\star)^{-4}$.
Eliminating $x_\star$ gives, at leading order
\begin{equation}
\label{nsSub}
  n_s\simeq1-\frac{4}{3\,N_\star}+\mathcal{O}\!\left(N_\star^{-4/3}\right),
\end{equation}
\begin{equation}
\label{rSub}
  r\simeq\mathcal{C}(\alpha,\beta)\,N_\star^{-4/3},
\end{equation}
where $\mathcal{C}(\alpha,\beta)=16\cdot3^{-5/3}\alpha^{-1/3}\beta^{-4/3}
\bigl(e^{-k}/(1-e^{-k})\bigr)^{2/3}$ and
$k=a/\beta=(1/\beta)\sqrt{2/(3\alpha)}\, ,$ (see Appendix).
Thus as $\beta$ increases from 0 the model flows from $r\propto N_\star^{-2}$ to $r\propto N_\star^{-4/3}$ and $n_s$ changes from
$1-2/N_\star$ to $1-4/(3N_\star)$.

\paragraph{(iii) Practical crossover criterion.}
The $\alpha$-Starobinsky-like behavior persists as long as the denominator $1+\beta x$ has little effect on the exponential tail over the observable $N_\star$ $e$-folds, i.e.
\begin{equation}
\label{crossoverCond}
  \beta\,x_\star\ll1.
\end{equation}
Since in $E$-models $x_\star\sim\mathcal{O}\!\bigl(\sqrt{\alpha}\ln N_\star\bigr)$, this translates to
\begin{equation}
\label{betaCriterion}
  \beta\ll\frac{1}{\sqrt{\alpha}\ln N_\star},
\end{equation}
in which case predictions remain very close to $\beta=0$. For larger values the dynamics progressively enters the submaximal regime.

\section{Implications for $n_s$, $r$, and $N_\star$}\label{Observables}

\begin{figure*}[h!]
\begin{center}
$\begin{array}{ccc}
\includegraphics[width=2.in]{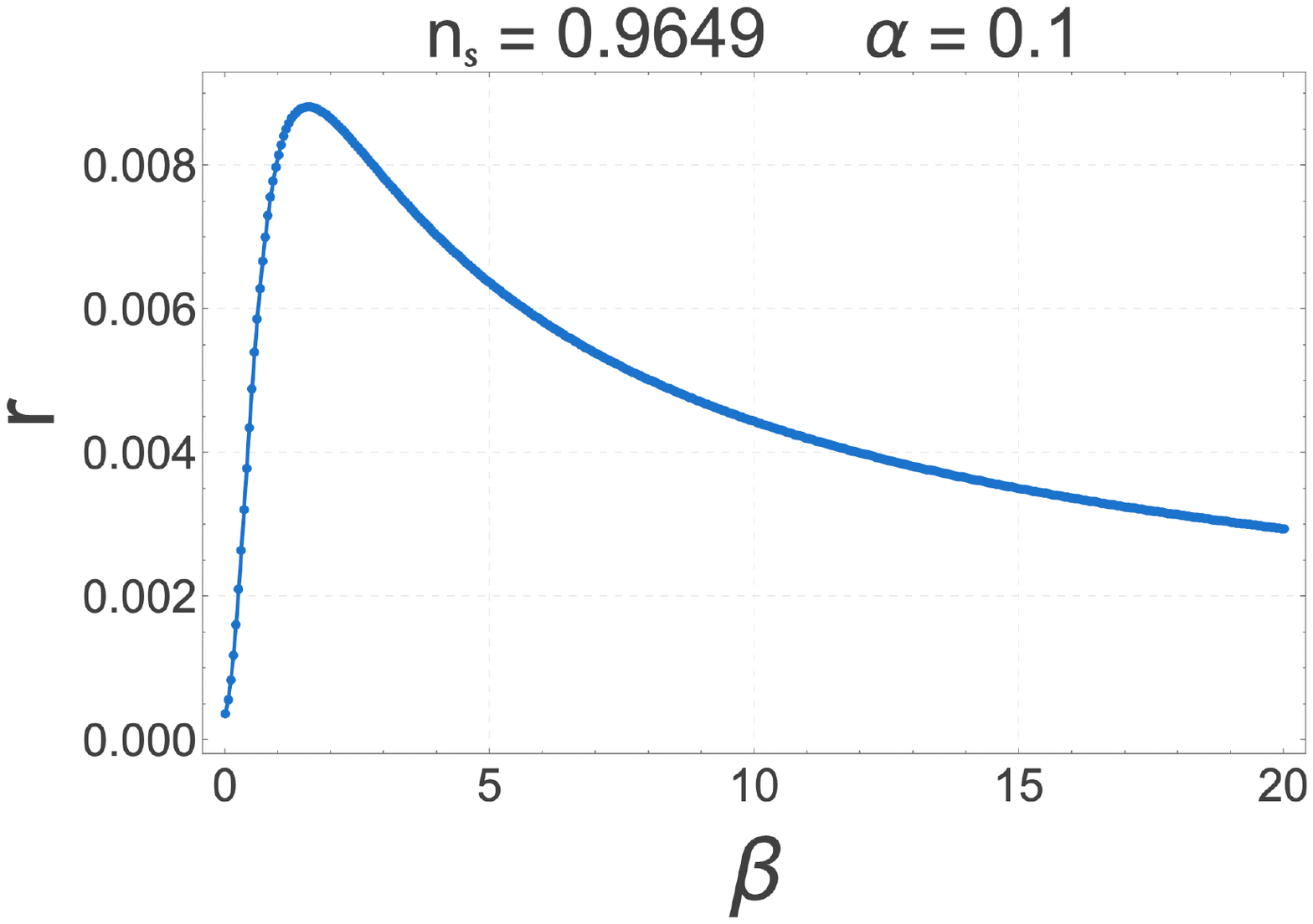}&
\includegraphics[width=2.in]{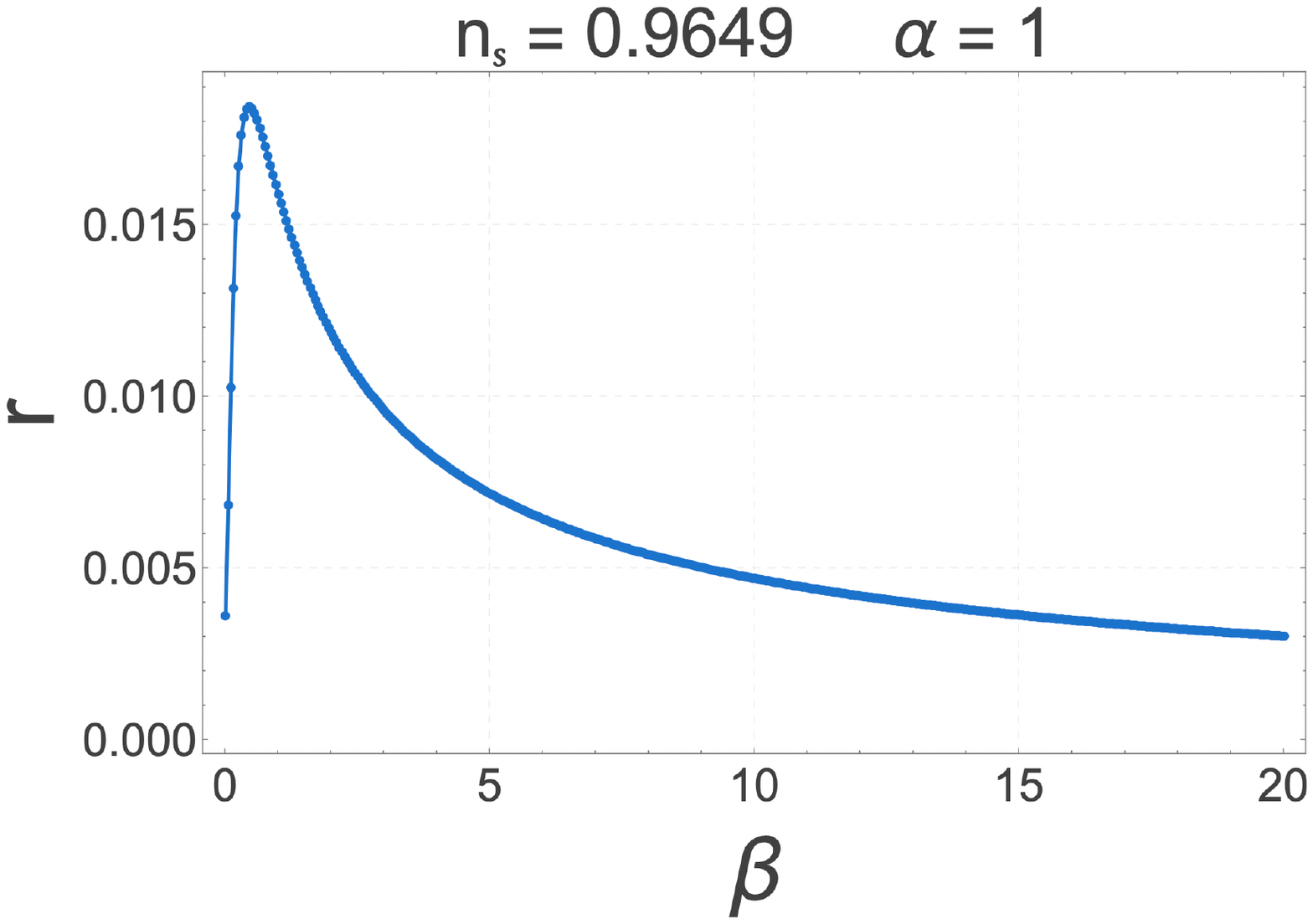}&
\includegraphics[width=2.in]{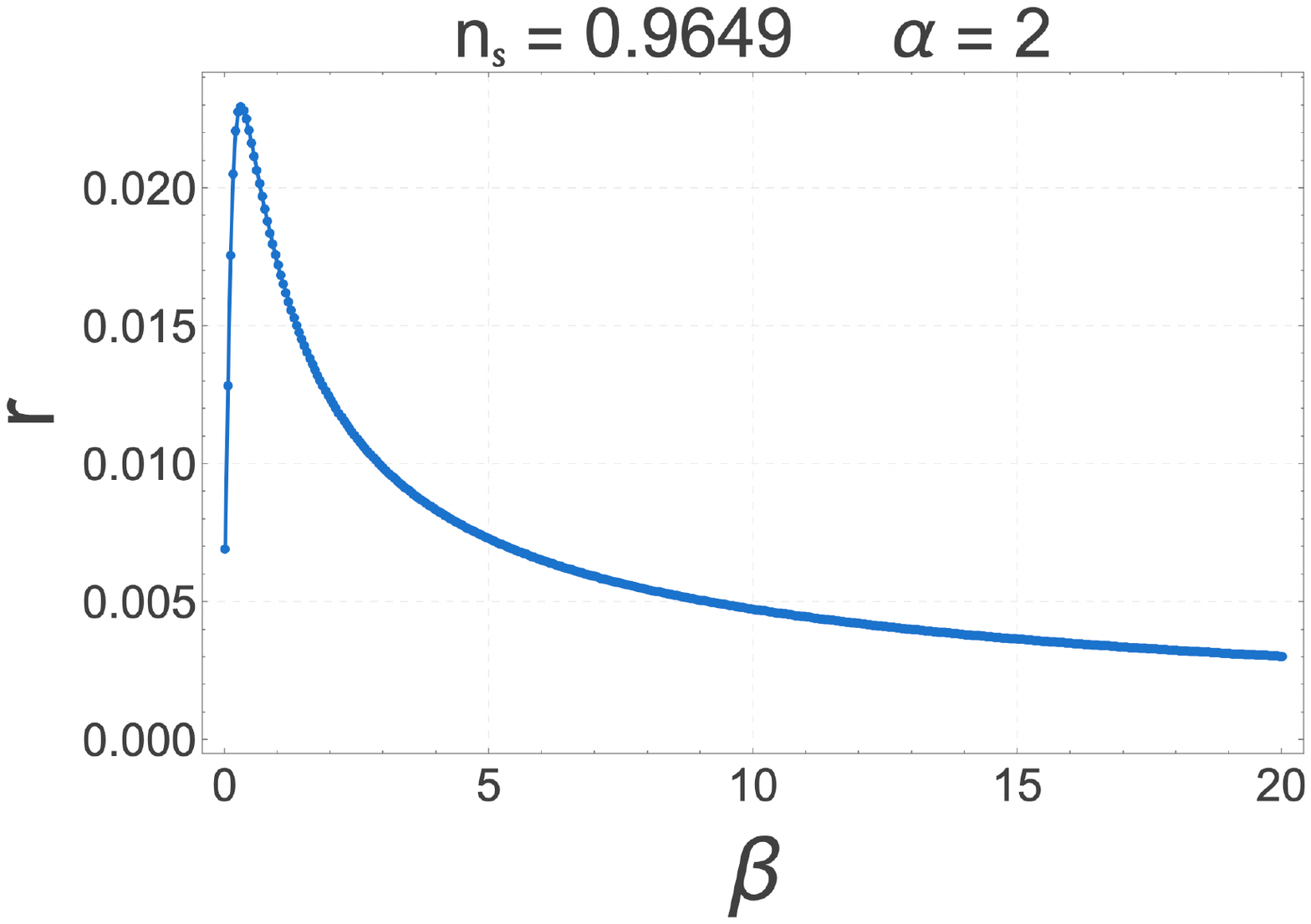}
\end{array}$
\caption{Tensor-to-scalar ratio $r$ as a function of the deformation parameter $\beta$ for the $\alpha\beta$-Starobinsky model defined by eq.~(\ref{ValphaBeta}), shown for three representative values of $\alpha$. The field value at horizon crossing $\phi_*$ is determined by fixing the spectral index to its central observed value, $n_s = 0.9649$, through the slow-roll relation $n_s = 1 + 2\eta - 6\epsilon$. The non-monotonic behavior of $r$ as a function of $\beta$ is also an interesting feature of the model \cite{German:2015qjq}.}
\label{figr}
\end{center}
\end{figure*}

\begin{figure*}[h!]
\begin{center}
$\begin{array}{ccc}
\includegraphics[width=2.in]{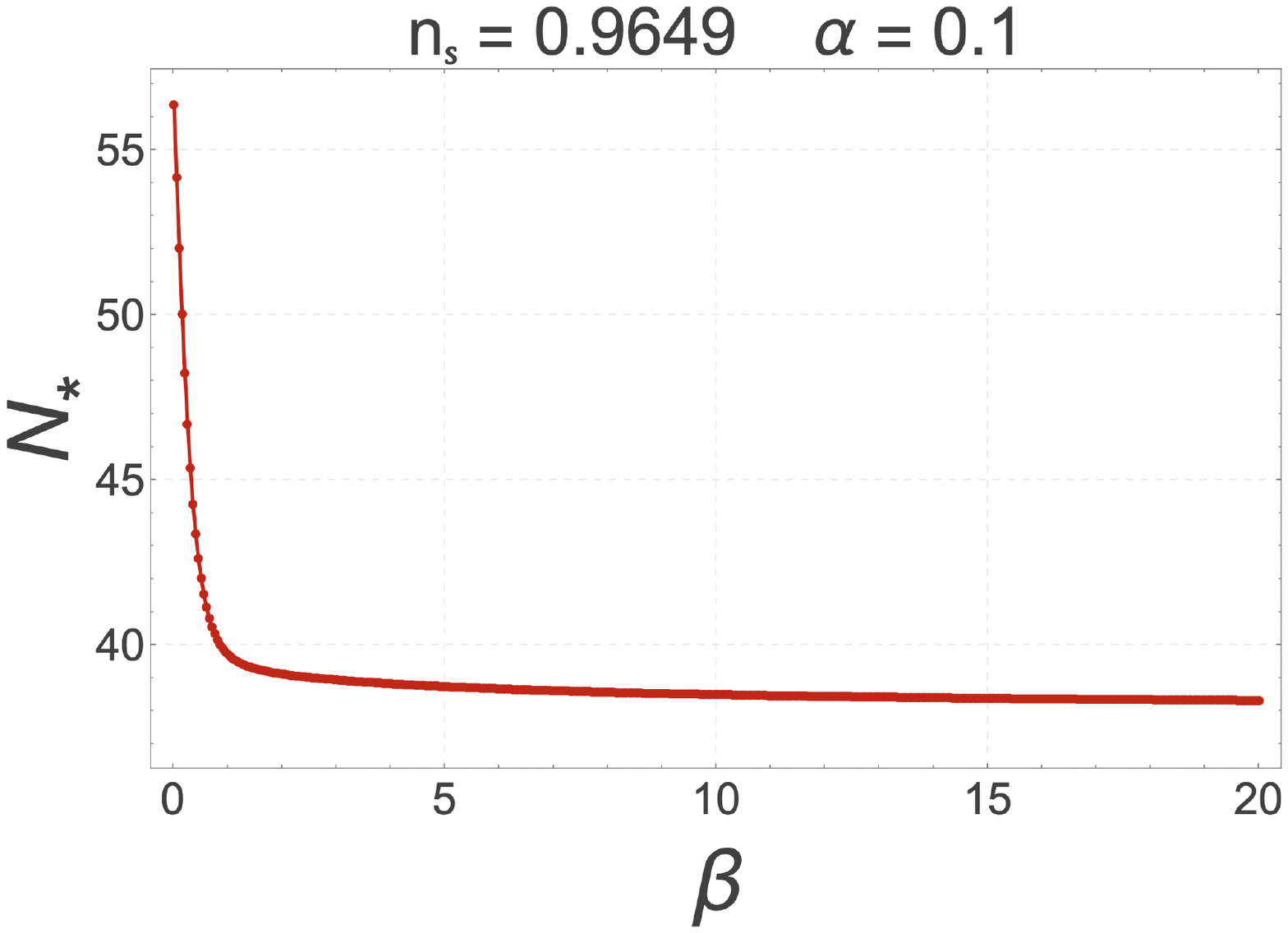}&
\includegraphics[width=2.in]{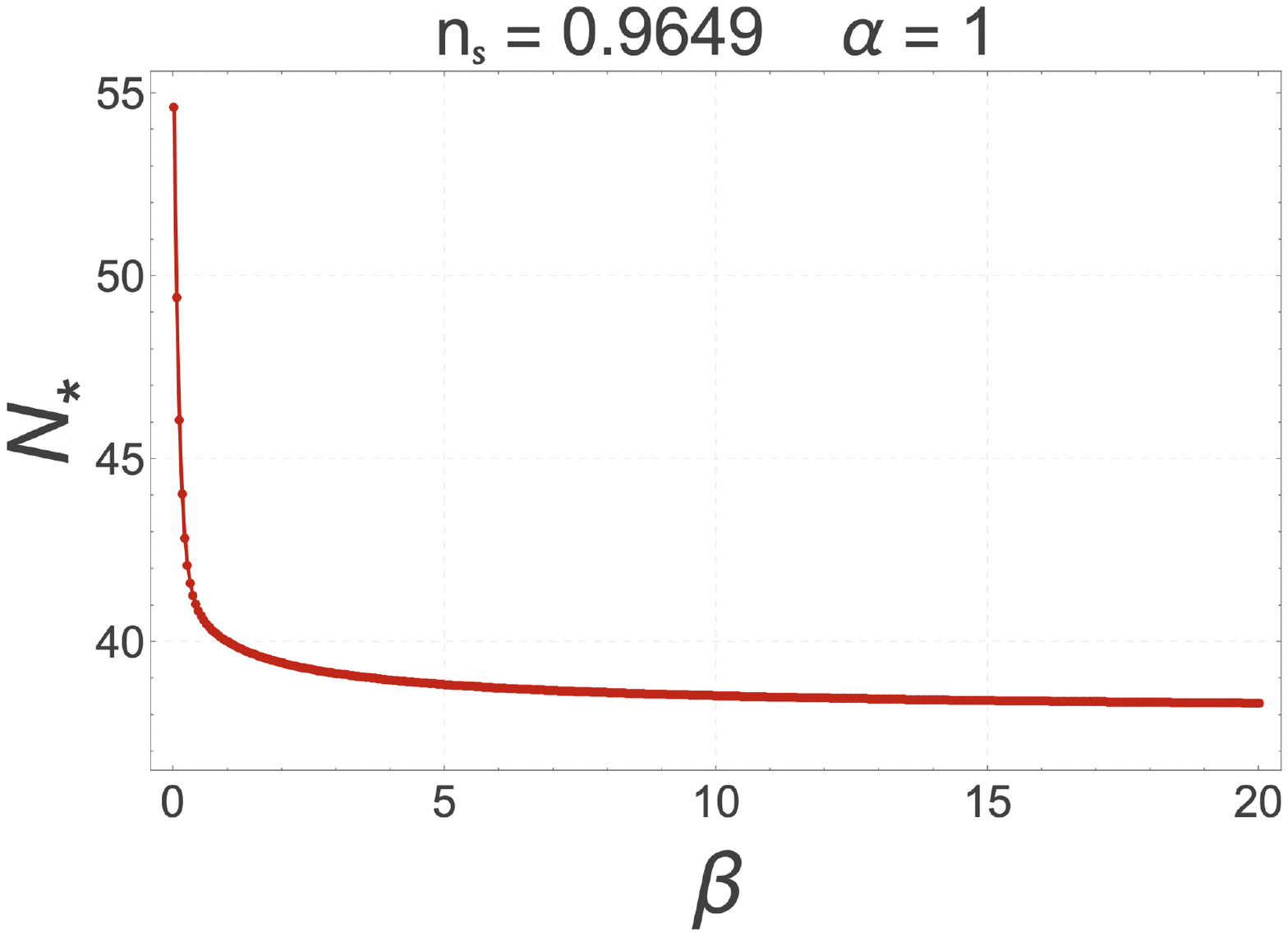}&
\includegraphics[width=2.in]{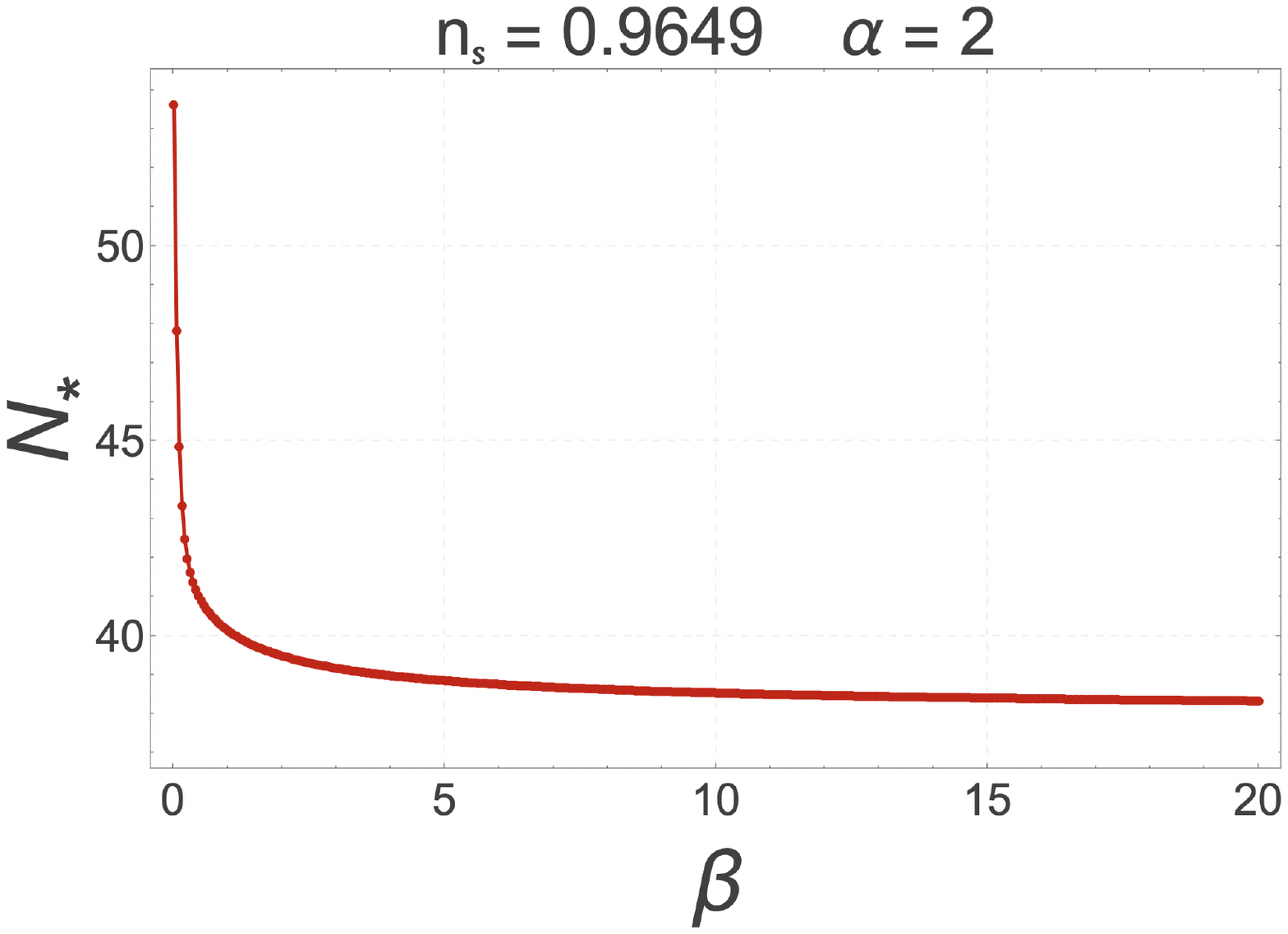}
\end{array}$
\caption{Number of $e$-folds $N_*$ accumulated between the horizon crossing of the pivot scale $k_*$ and the end of inflation, as a function of the deformation parameter $\beta$ for the $\alpha\beta$-Starobinsky model defined by eq.~(\ref{ValphaBeta}), shown for the same three values of $\alpha$ as in Fig.~\ref{figr}. As before, $\phi_*$ is determined by imposing $n_s = 0.9649$ via the slow-roll relation $n_s = 1 + 2\eta - 6\epsilon$.}
\label{N*0.9649}
\end{center}
\end{figure*}

\begin{figure*}[h!]
\begin{center}
$\begin{array}{ccc}
\includegraphics[width=2.in]{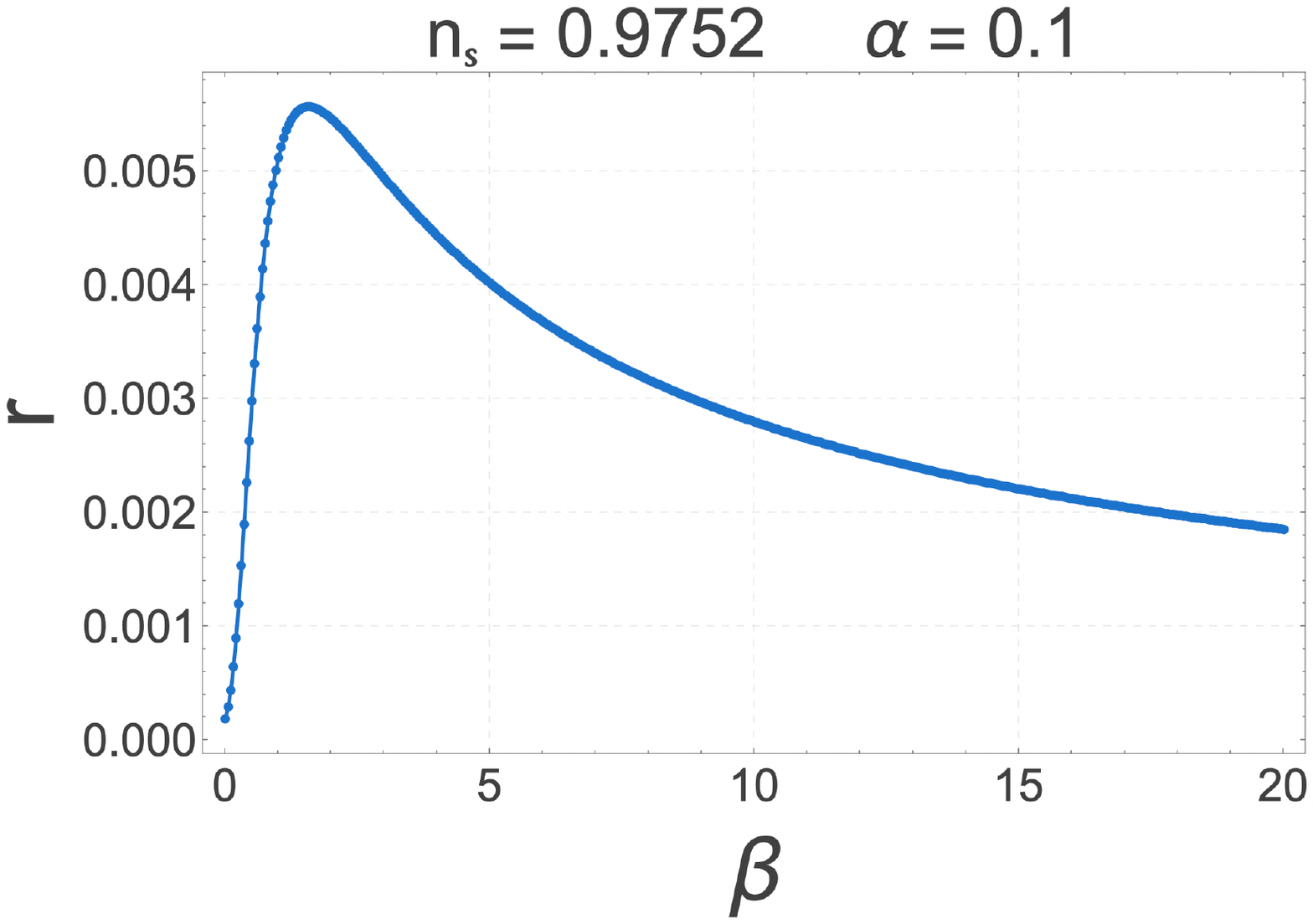}&
\includegraphics[width=2.in]{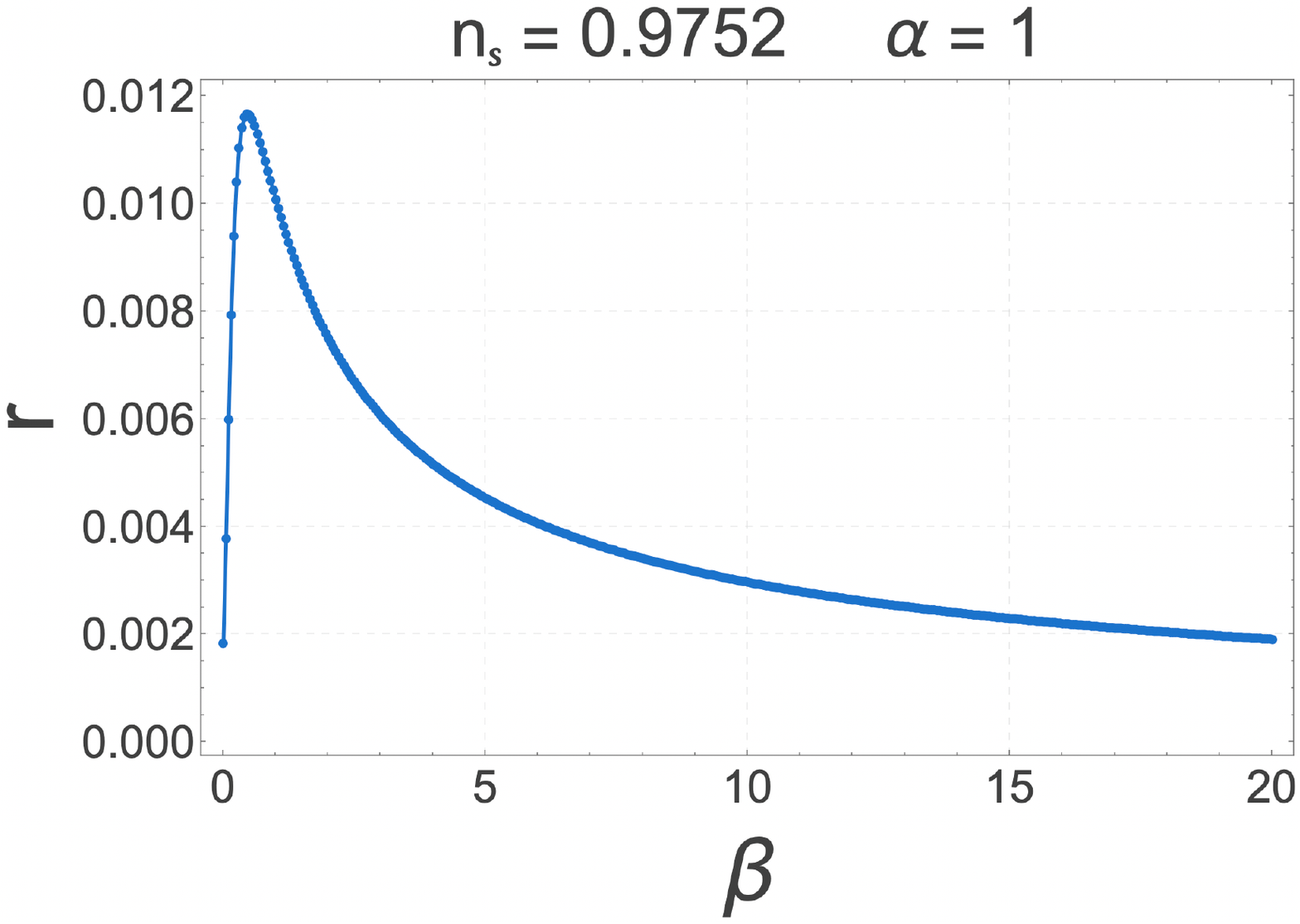}&
\includegraphics[width=2.in]{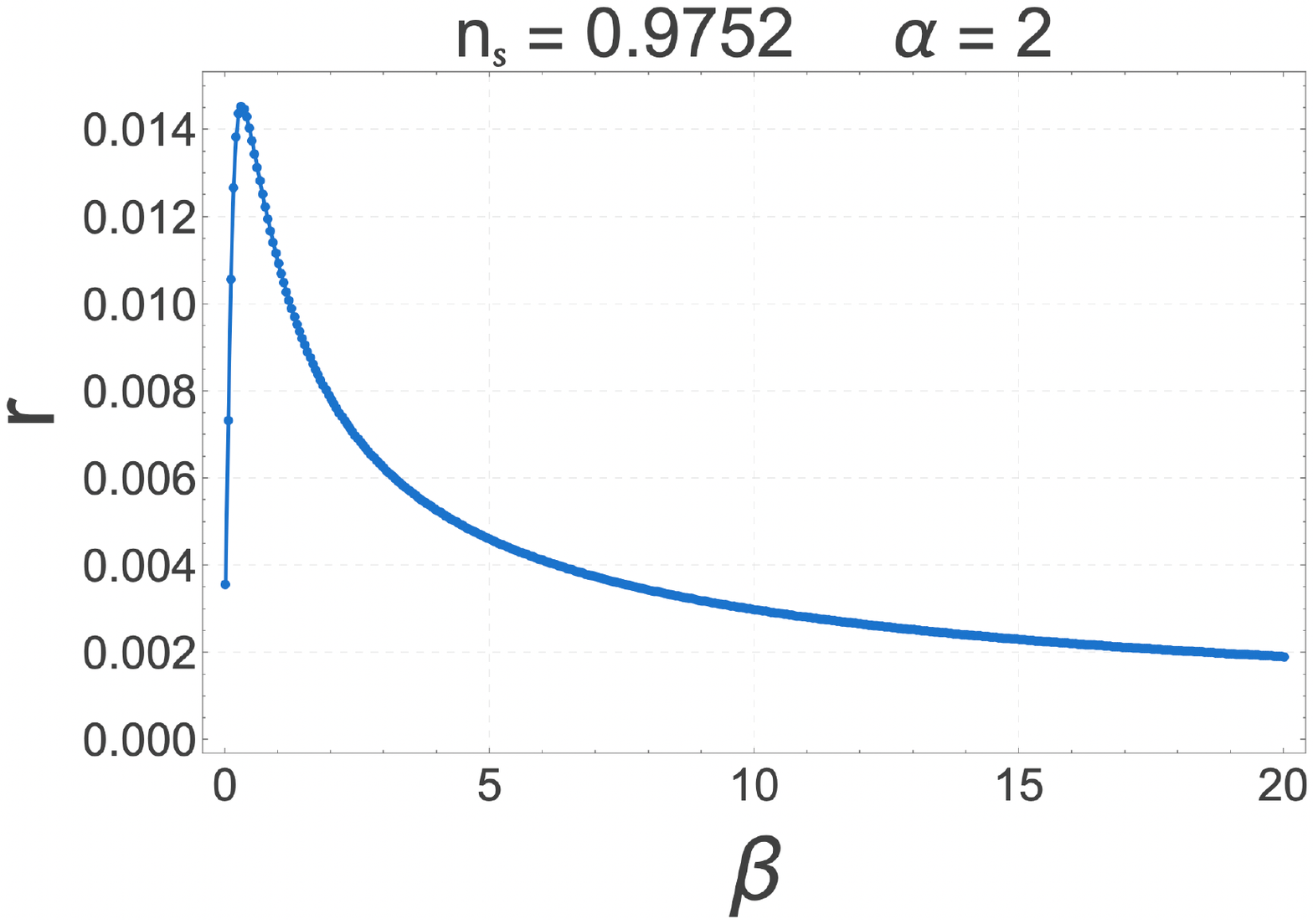}
\end{array}$
\caption{As in Fig.~\ref{figr} with the spectral index fixed at $n_s = 0.9752.$}
\label{figr0.9752}
\end{center}
\end{figure*}

\begin{figure*}[h!]
\begin{center}
$\begin{array}{ccc}
\includegraphics[width=2.in]{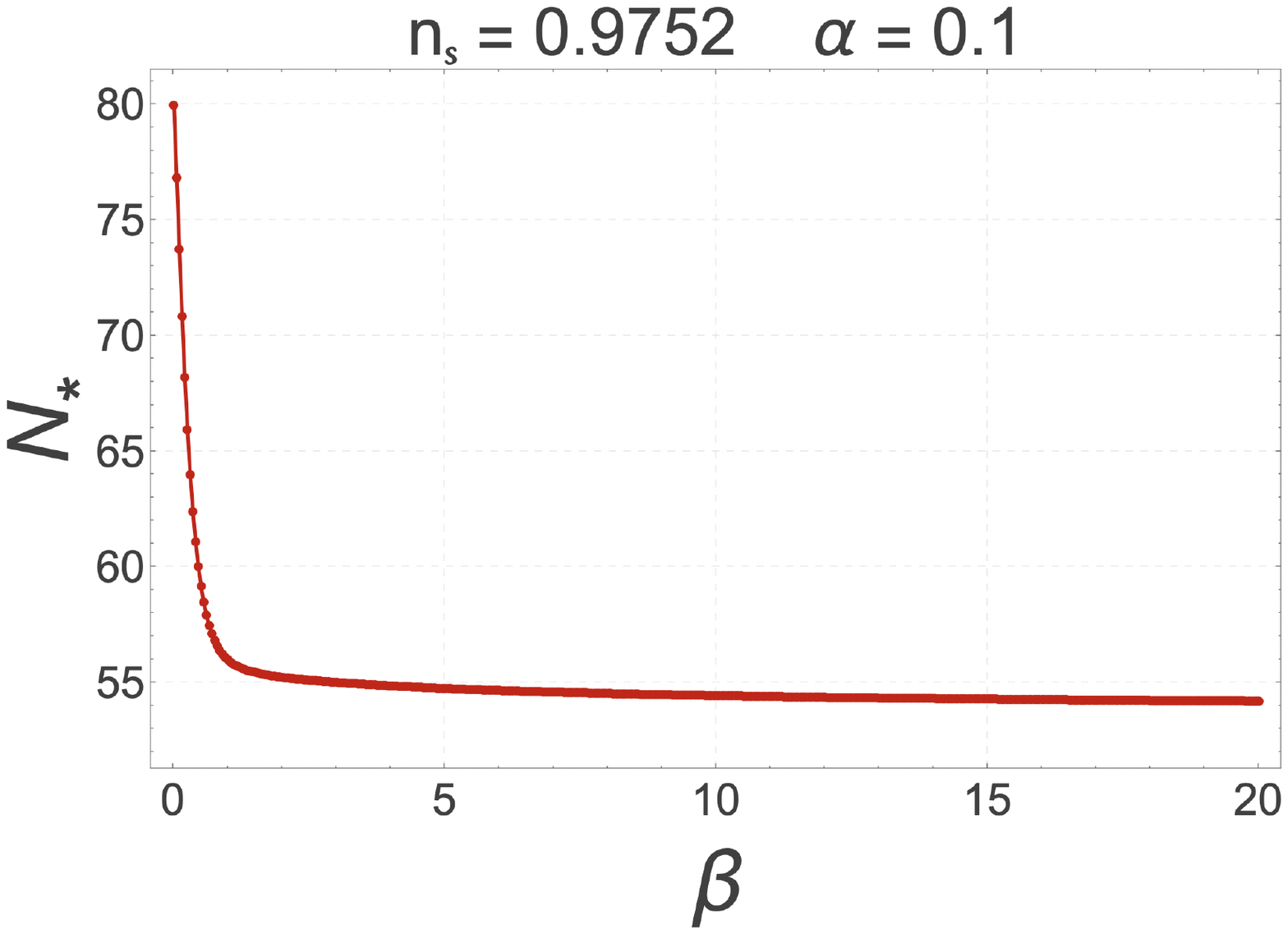}&
\includegraphics[width=2.in]{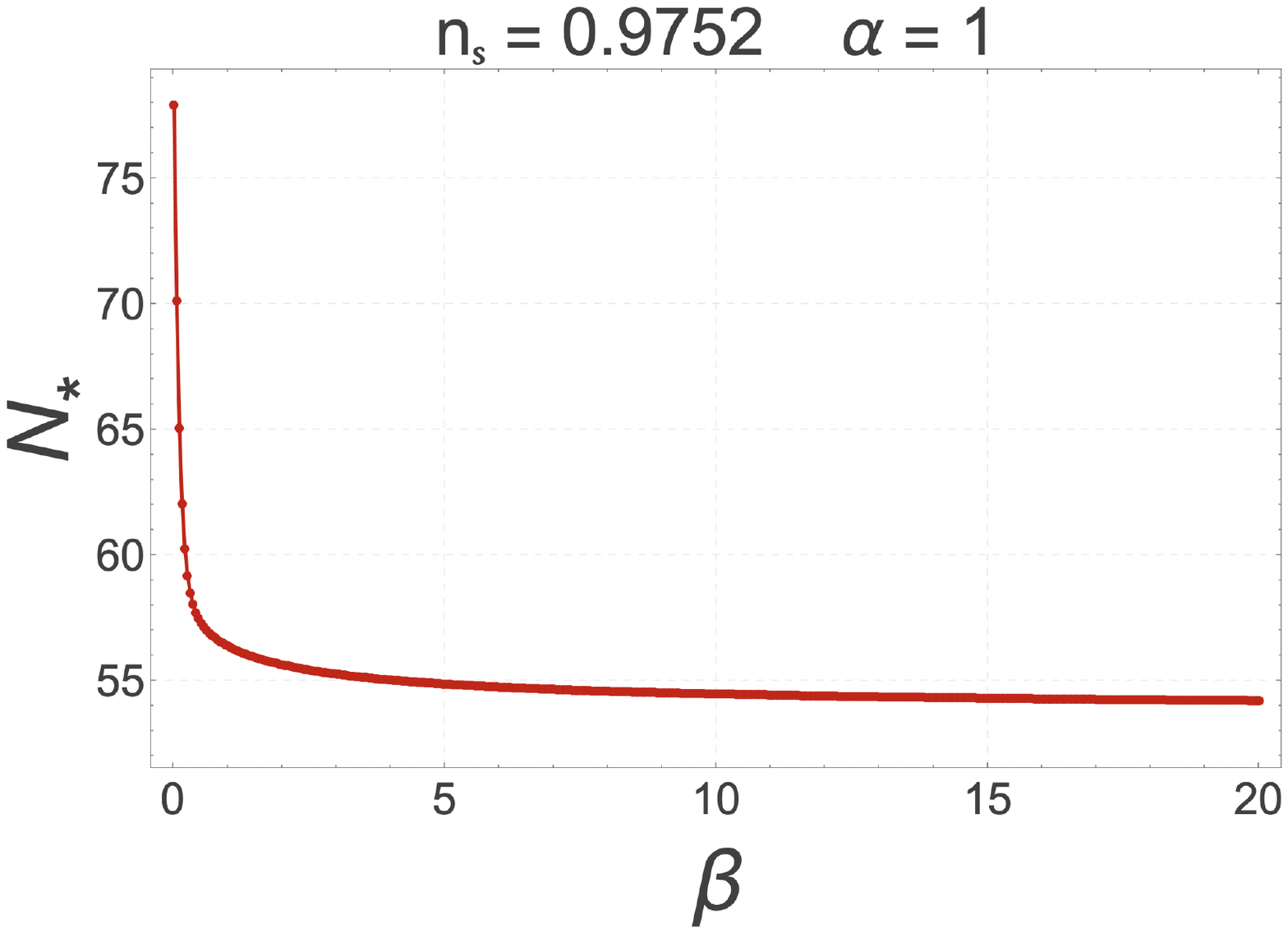}&
\includegraphics[width=2.in]{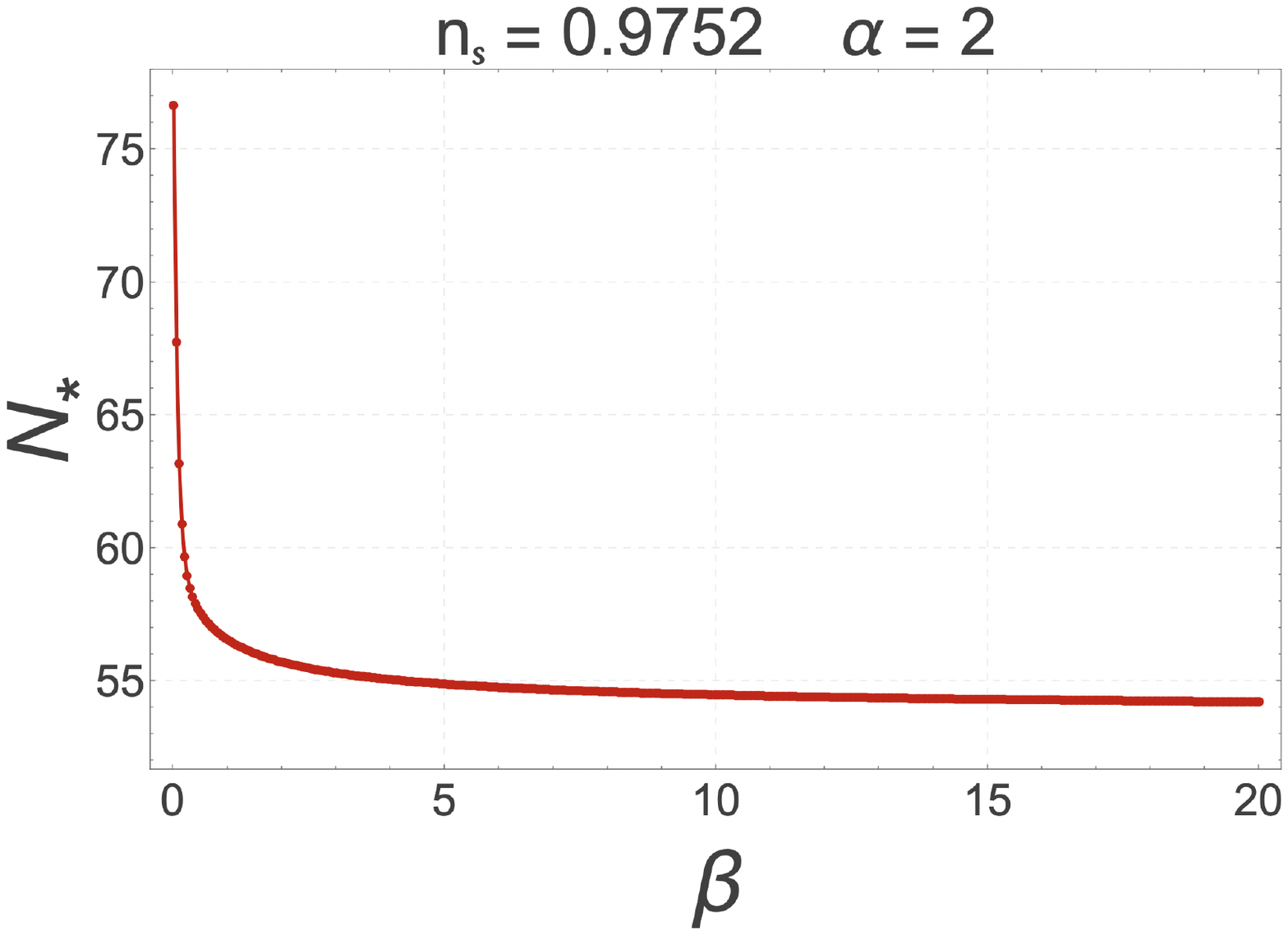}
\end{array}$
\caption{As in Fig.~\ref{N*0.9649} with the spectral index fixed at $n_s = 0.9752.$}
\label{N*0.9752}
\end{center}
\end{figure*}

The parameter $\alpha$ sets the effective curvature scale (in the $\beta=0$ limit) and controls $r$ via eq.~\eqref{alphaRegime}, while $\beta>0$ truncates the plateau to a height $V_\infty<V_0$ and changes the scaling laws to eqs.~\eqref{nsSub}--\eqref{rSub}. For a fixed observed $n_s$ one infers
\begin{equation}
\label{NstarFromns}
  N_\star\simeq
  \begin{cases}
    \dfrac{2}{1-n_s},         & \text{$\alpha$-Starobinsky regime, $\beta\!\to\!0$},\\[8pt]
    \dfrac{4}{3(1-n_s)}, & \text{$\alpha\beta$-Starobinsky regime, $\beta>0$ sufficiently large},
  \end{cases}
\end{equation}
with $r$ then following from eq.~\eqref{alphaRegime} or eq.~\eqref{rSub}, respectively. Figures \ref{figr} to \ref{N*0.9752} show the behavior of the tensor-to-scalar ratio $r$ and the number of $e$-folds $N_*$ as functions of the parameter $\beta$ for three characteristic values of $\alpha$ and Table~\ref{numericalvalues} give a representative sample of cosmological observables values for the $\alpha\beta$-Starobinsky model defined by eq.~(\ref{ValphaBeta}).

\newpage

\begin{table*}[htbp!]
\begin{center}
\begin{tabular}{ccccc}
Parameters &\,\,$n_{sk}$   &\,\,   $r$   &\,\, $N_{*}$   &\,\, $\Delta (GeV)$\\ \hline
& \quad   & \quad  \\[-2mm]
$\alpha=1$,\,\,$\beta=0,$\,\,$n_s=0.9649$ & \quad $-6.2\times 10^{-4}$ & \quad $0.0035$  & \quad $54.9$ & \quad $4.4\times 10^{15}$ \\[2mm]
$\alpha=1$,\,\,$\beta=0,$\,\,$n_s=0.9752$ & \quad $-3.1\times 10^{-4}$ & \quad $0.0018$ & \quad $78.3$ & \quad $3.7\times 10^{15}$ \\[2mm]
$\alpha=1$,\,\,$\beta=2,$\,\,$n_s=0.9649$ & \quad $-9.0\times 10^{-4}$ & \quad $0.0119$  & \quad $39.4$ & \quad $6.0\times 10^{15}$ \\[2mm]
$\alpha=1$,\,\,$\beta=2,$\,\,$n_s=0.9752$  & \quad $-4.5\times 10^{-4}$  & \quad $0.0075$ & \quad $55.6$ & \quad $5.4\times 10^{15}$ \\[2mm]
\end{tabular}
\caption{\label{parameters} A representative sample of cosmological observables is tabulated for the $\alpha\beta$-Starobinsky model defined by eq.~(\ref{ValphaBeta}). The first two rows correspond to the standard Starobinsky model ($\beta = 0$) and serve as a baseline for comparison. Notably, the higher spectral index $n_s = 0.9752$ yields an unusually large number of $e$-folds ($N_* = 78.3$), well outside the conventionally preferred range, whereas the $\alpha\beta$-Starobinsky model with $\beta = 2$ brings $N_*$ comfortably within the expected interval $N_*\in(50,60)$. The running of the scalar spectral index $n_s$ is here denoted by $n_{sk}.$}
\label{numericalvalues}
\end{center}
\end{table*}

\section{Conclusions}\label{Con}

The $\alpha\beta$-Starobinsky model, eq.~\eqref{ValphaBeta}, provides a continuous bridge between a \emph{maximal} plateau ($\beta\!\to\!0$) and a \emph{submaximal} plateau ($\beta>0$), preserving single-field slow-roll simplicity while altering the asymptotic approach to the plateau. The key phenomenological consequence, summarized by eqs.~\eqref{alphaRegime}--\eqref{rSub}, is a continuous flow of $(n_s,r)$ predictions that lifts $n_s$ at fixed $N_\star$ and softens the $N_\star$-scaling of $r$ from $N_\star^{-2}$ to $N_\star^{-4/3}$.
This enables excellent agreement with Planck~2018 constraints and accommodation of the higher $n_s$ preferred by ACT+DESI~DR2 (BAO), within the theoretically motivated interval $N_\star\in(50,60)$, in contrast to standard Starobinsky and $\alpha$-Starobinsky models, which typically require $N_\star\gtrsim60$ or tailored reheating.

On the theoretical side, the paper provides an exact integral for $N_\star$, eq.~\eqref{Nstary}, facilitating controlled approximations, the crossover criterion $\beta x_\star\lessgtr1$, eq.~\eqref{betaCriterion}, and closed-form predictions for $(n_s,r)$ in both regimes, eqs.~\eqref{alphaRegime} and \eqref{nsSub}, \eqref{rSub}. Together these make it straightforward to map future $(n_s,r)$ measurements into constraints on $(\alpha,\beta)$. The distinctive softened scaling $r\propto N_\star^{-4/3}$ is directly testable with
short-term CMB polarization and BAO improvements.

Natural next steps include a full joint likelihood analysis with Planck, ACT, SPT, BICEP/Keck, and DESI~DR2, forecasts for LiteBIRD sensitivity to the $(\alpha,\beta)$ flow, exploration of reheating microphysics consistent with the favored parameter space. Taken together, these elements will make it possible to refine the observational distinction between maximal and submaximal inflationary plateaus and could determine the degree of deformation that occurred in the early universe.

\section*{Acknowledgements}

I would like to thank DGAPA-PAPIIT-UNAM grant No. IN110325 {\it Estudios en cosmolog\'ia inflacionaria, agujeros negros primordiales y energ\'ia oscura}.

\appendix

\section{Details of the continuous crossover}
\label{app:A}

Define 
\begin{equation}
\label{a1}
E(x)\equiv e^{y(x)}=\exp\!\left(-\,\frac{a\,x}{1+\beta x}\right),
\qquad
a^2=\frac{2}{3\alpha}, \quad a>0,
\qquad
k\equiv \frac{a}{\beta}.
\end{equation}
In the asymptotic ``plateau'' regime relevant for CMB scales, $y(x)$ approaches a constant $y\to -k$, so
\begin{equation}
\label{a2}
E(x)\approx E_\infty \equiv e^{-k}
\quad \text{(slowly varying, treat as constant to leading order).}
\end{equation}
From the given expressions,
\begin{equation}
\label{a3}
\frac{V'}{V}=\frac{2a}{M_{\mathrm{Pl}}}\,\frac{E}{1-E}\,\frac{1}{(1+\beta x)^{2}},
\end{equation}
hence
\begin{equation}
\label{a4}
\epsilon
=\frac{1}{2}M_{\mathrm{Pl}}^{2}\!\left(\frac{V'}{V}\right)^{2}
=\frac{4}{3\alpha}\left(\frac{E}{1-E}\right)^{2}\frac{1}{(1+\beta x)^{4}}.
\end{equation}
In the plateau limit $E\approx E_\infty$ is constant, so the only $x$-dependence is
\begin{equation}
\label{a5}
\epsilon(x)\;\propto\;(1+\beta x)^{-4}
\quad \text{(with constant prefactor).}
\end{equation}

The $e$-fold integrand is
\begin{equation}
\label{a6}
\frac{V}{V'}=\frac{M_{\mathrm{Pl}}}{2a}\,\frac{1-E}{E}\,(1+\beta x)^{2}.
\end{equation}
With $x=\phi/M_{\mathrm{Pl}}$ (so $d\phi/M_{\mathrm{Pl}}=dx$), the $e$-fold number
from $x_{\rm end}$ to $x_\star$ is
\begin{equation}
\label{a7}
N_\star=\int_{x_{\rm end}}^{x_\star}\frac{V}{V'}\,\frac{d\phi}{M_{\mathrm{Pl}}}
=\frac{1}{2a}\int_{x_{\rm end}}^{x_\star}
\frac{1-E(x)}{E(x)}\,(1+\beta x)^{2}\,dx.
\end{equation}
In the plateau regime take $\frac{1-E}{E}\approx \frac{1-E_\infty}{E_\infty}$ as
a constant:
\begin{equation}
\label{a8}
N_\star \;\simeq\; \frac{1}{2a}\,\frac{1-E_\infty}{E_\infty}
\int_{x_{\rm end}}^{x_\star}(1+\beta x)^{2}\,dx
= \frac{1}{2a}\,\frac{1-E_\infty}{E_\infty}\;
\frac{(1+\beta x)^{3}}{3\beta}\Bigg|_{x_{\rm end}}^{x_\star}.
\end{equation}
Provided $(1+\beta x_\star)^3 \gg (1+\beta x_{\rm end})^3$, the lower limit is
negligible and
\begin{equation}
\label{a9}
N_\star \;\simeq\; A\,(1+\beta x_\star)^{3}\ ,\qquad
A \equiv \frac{1}{6a\beta}\,\frac{1-E_\infty}{E_\infty}.
\end{equation}
This is the announced cubic scaling $N_\star \propto (1+\beta x_\star)^3$. Using the plateau approximation in $\epsilon$,
\begin{equation}
\label{a10}
\epsilon_{\star} \;\simeq\; B\,(1+\beta x_\star)^{-4},\qquad
B \equiv \frac{4}{3\alpha}\left(\frac{E_\infty}{1-E_\infty}\right)^{2}.
\end{equation}
From the cubic law,
\begin{equation}
\label{a11}
(1+\beta x_\star) \;\simeq\; \left(\frac{N_\star}{A}\right)^{1/3}
\quad\Longrightarrow\quad
\epsilon_{\star} \;\simeq\; B\,\left(\frac{N_\star}{A}\right)^{-4/3}
= B\,A^{4/3}\,N_\star^{-4/3}.
\end{equation}
Therefore
\begin{equation}
\label{a12}
r \equiv 16\epsilon_{\star} \;\simeq\;
\mathcal{C}(\alpha,\beta)\,N_\star^{-4/3}\ ,\qquad
\mathcal{C}(\alpha,\beta)=16\,B\,A^{4/3}.
\end{equation}
Insert $A$ and $B$ and use $a^2=2/(3\alpha)$, $E_\infty=e^{-k}$, $k=a/\beta$:
\begin{equation}
\label{a13}
\begin{aligned}
\mathcal{C}(\alpha,\beta)
&= 16\;\frac{4}{3\alpha}\left(\frac{E_\infty}{1-E_\infty}\right)^{2}
\left[\frac{1}{6a\beta}\,\frac{1-E_\infty}{E_\infty}\right]^{\!4/3} \\
&= 16\cdot \frac{4}{3\alpha}\cdot \frac{E_\infty^{2}}{(1-E_\infty)^{2}}\cdot
(6a\beta)^{-4/3}\cdot \left(\frac{1-E_\infty}{E_\infty}\right)^{4/3} \\
&= 16\cdot \frac{4}{3\alpha}\cdot (6a\beta)^{-4/3}\cdot
\left(\frac{E_\infty}{1-E_\infty}\right)^{2/3}.
\end{aligned}
\end{equation}
Collecting numeric and $\alpha$ powers gives
\begin{equation}
\label{a14}
\mathcal{C}(\alpha,\beta)
=16\,3^{-5/3}\,\alpha^{-1/3}\,\beta^{-4/3}
\left(\frac{e^{-k}}{1-e^{-k}}\right)^{2/3}.
\end{equation}
Using the potential slow-roll relation
\begin{equation}
\label{a15}
n_s - 1 \;=\; -6\epsilon + 2\eta,
\qquad
\eta \equiv M_{\mathrm{Pl}}^{2}\frac{V''}{V}
= M_{\mathrm{Pl}}^{2}\frac{d}{d\phi}\!\left(\frac{V'}{V}\right)+2\epsilon,
\end{equation}
we get
\begin{equation}
\label{a16}
n_s - 1 \;=\; -2\epsilon + 2 M_{\mathrm{Pl}}^{2}
\frac{d}{d\phi}\!\left(\frac{V'}{V}\right).
\end{equation}
In the plateau limit $E\approx E_\infty$ is constant, so
\begin{equation}
\label{a17}
\frac{V'}{V}\;\approx\;
\frac{2a}{M_{\mathrm{Pl}}}\frac{E_\infty}{1-E_\infty}\,(1+\beta x)^{-2}
\;\equiv\; C_0\,(1+\beta x)^{-2}, \qquad
C_0=\frac{2a}{M_{\mathrm{Pl}}}\frac{E_\infty}{1-E_\infty}.
\end{equation}
Hence, using $d/d\phi=(1/M_{\mathrm{Pl}})\,d/dx$,
\begin{equation}
\label{a18}
M_{\mathrm{Pl}}^{2}\frac{d}{d\phi}\!\left(\frac{V'}{V}\right)
= M_{\mathrm{Pl}}^{2}\cdot \frac{1}{M_{\mathrm{Pl}}}\cdot
\frac{d}{dx}\!\left(C_0(1+\beta x)^{-2}\right)
= -\,2 C_0\,\beta\,M_{\mathrm{Pl}}\,(1+\beta x)^{-3}.
\end{equation}
Therefore, up to subleading
$\mathcal{O}(\epsilon)\sim \mathcal{O}(N_\star^{-4/3})$,
\begin{equation}
\label{a19}
n_s - 1 \;\simeq\;
2\bigl[-\,2 C_0\,\beta\,M_{\mathrm{Pl}}\,(1+\beta x_\star)^{-3}\bigr]
= -\,8 a \beta \frac{E_\infty}{1-E_\infty}\,(1+\beta x_\star)^{-3}.
\end{equation}
Use the cubic law to eliminate $x_\star$:
\begin{equation}
\label{a20}
(1+\beta x_\star)^{-3} \;\simeq\; \frac{A}{N_\star}
= \frac{1}{N_\star}\,\frac{1}{6a\beta}\,\frac{1-E_\infty}{E_\infty}.
\end{equation}
This gives the leading-order tilt
\begin{equation}
\label{a21}
n_s \;\simeq\; 1 - \frac{4}{3\,N_\star}
\;+\; \mathcal{O}\!\bigl(N_\star^{-4/3}\bigr).
\end{equation}
The $\mathcal{O}(N_\star^{-4/3})$ corrections arise from the subleading $-2\epsilon$ term and from the mild $x$-dependence of $E(x)$ neglected at leading order. Thus, as $\beta$ is dialed from $0$ to $\beta>0$, one flows from the usual Starobinsky-type behavior $r\propto N_\star^{-2}$, $n_s\simeq 1-\frac{2}{N_\star}$ to the modified scalings $r\propto N_\star^{-4/3}$, $n_s\simeq 1-\frac{4}{3N_\star}$.

\end{document}